\documentclass[conference,table]{IEEEtran}
\IEEEoverridecommandlockouts

\usepackage[named]{algo}
\usepackage{amssymb}
\usepackage{amsfonts}
\usepackage{amsmath}
\usepackage{algorithm}
\usepackage{algorithmic}
\usepackage{ulem}
\usepackage{color}
\usepackage{verbatim}
\usepackage{multirow}
\usepackage{graphicx}
\usepackage{graphics}
\usepackage{balance}
\usepackage{mathtools}
\usepackage{pgfplots}
\usepackage{caption}
\usepackage{subcaption}
\usepackage{rotating}
\usepackage{highlight}

\newcommand{\Comment}[1]{}

\usepackage[named]{algo}
\usepackage{xspace}
\usepackage{algorithm}
\usepackage{algorithmic}

\def\x{{\mathbf x}}

\def\R{\rm I\!R}

\providecommand{\abs}[1]{\lvert#1\rvert}

\definecolor{light}{gray}{.9}

\newcommand{\offset}{\textsc{Offset}\xspace}

\newcommand{\argmax}{\operatornamewithlimits{argmax}}
\newcommand{\argmin}{\operatornamewithlimits{argmin}}

\newcommand{\oren}[1]{}

\newcommand{\alex}[1]{}

\usepackage{etoolbox} \makeatletter \patchcmd{\maketitle}{\@copyrightspace}{}{}{} \makeatother

\begin{document}

\Comment{
To do list:
1. Add Rina's paper
2. why it is not compared to CTR based DCO
}

\title{Conversion-Based Dynamic-Creative-Optimization \\ in Native Advertising}

\Comment{
\author{Yohay Kaplan}
\affiliation{Yahoo Research, Haifa, Israel}
\email{yohay@verizonmedia.com}

\author{Yair Koren}
\affiliation{Yahoo Research, Haifa, Israel}
\email{yairkoren@verizonmedia.com}

\author{Alex Shtoff}
\affiliation{Yahoo Research, Haifa, Israel}
\email{alex.shtoff@verizonmedia.com}

\author{Tomer Shadi}
\affiliation{Yahoo Product, Ramat-Gan, Israel}
\email{tomer.shadi@verizonmedia.com}

\author{Oren Somekh}
\affiliation{Yahoo Research, Haifa, Israel}
\email{orens@verizonmedia.com}
}

\Comment{
\author{Yohay Kaplan, Yair Koren, Alex Shtoff, Oren~Somekh}
\affiliation{Yahoo Research, Haifa, Israel}
\email{{yohay,yairkoren,alex.shtoff,orens}@verizonmedia.com}

\author{Tomer Shadi}
\affiliation{Yahoo Product, Ramat-Gan, Israel}
\email{tomer.shadi@verizonmedia.com}
}
\Comment{
\author{Yohay Kaplan, Yair Koren, Alex Shtoff, Tomer Shadi, and Oren~Somekh}
\affiliation{Yahoo Research, Haifa, Israel}
\email{{yohay,yairkoren,alex.shtoff,tomer.shadi,orens}@yahooinc.com}
}

%\Comment{
\author{\IEEEauthorblockN{Yohay Kaplan, Yair Koren, Alex Shtoff, Tomer Shadi, and Oren Somekh}
\IEEEauthorblockA{\textit{Yahoo Research} \\
Haifa, Israel\\
\{yohay,yairkoren,alex.shtoff,tomer.shadi,orens\}@yahooinc.com}}
%}
%\IEEEpubid{0000--0000/00\$00.00˜\copyright˜2022 IEEE}
\IEEEpubid{\copyright˜2022 IEEE}

%\author{\IEEEauthorblockN{Anonymous}}

\maketitle

\begin{abstract}
Yahoo Gemini native advertising marketplace serves billions of impressions daily, to hundreds millions of unique users, and reaches a yearly revenue of many hundreds of millions USDs. Powering Gemini native models for predicting advertise (ad) event probabilities, such as conversions and clicks, is \offset\ - a feature enhanced \textit{collaborative-filtering} (CF) based event prediction algorithm. The predicted probabilities are then used in Gemini native auctions to determine which ads to present for every serving event (impression). \textit{Dynamic creative optimization} (DCO) is a recent Gemini native product that was launched two years ago and is increasingly gaining more attention from advertisers. The DCO product enables advertisers to issue several assets per each native ad attribute, creating multiple combinations for each DCO ad. Since different combinations may appeal to different crowds, it may be beneficial to present certain combinations more frequently than others to maximize revenue while keeping advertisers and users satisfied. The initial DCO offer was to optimize \textit{click-through rates} (CTR), however as the marketplace shifts more towards conversion based campaigns, advertisers also ask for a \textit{conversion based solution}. To accommodate this request, we present a post-auction solution, where DCO ads combinations are favored according to their predicted \textit{conversion rate} (CVR). The predictions are provided by an \textit{auxiliary} \offset based combination CVR prediction model, and used to generate the combination distributions for DCO ad rendering during serving time. An online evaluation of this \textit{explore-exploit} solution, via online bucket A/B testing, serving Gemini native DCO traffic, showed a 53.5\% CVR lift, when compared to a control bucket serving all combinations uniformly at random. The impressive results demonstrate the ability of this practical yet effective solution to overcome many real-life issues such as data sparsity, reporting delays, trends, and various system constraints. The CVR prediction based DCO product is now fully deployed, serving all Gemini native traffic. 
\end{abstract}

\Comment{
\begin{figure}[t]
\centering
\includegraphics[width=0.6\columnwidth]{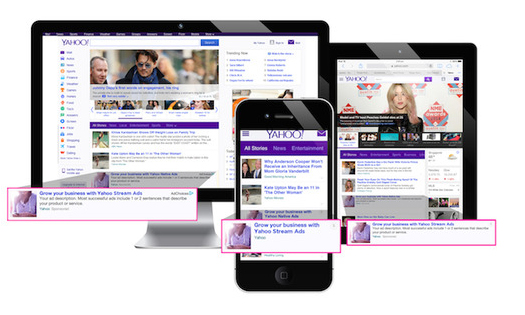}
\caption{Verizon Media native ads on different devices.}
\label{fig:Gemini native}
\end{figure}
}

\section{Introduction}\label{sec: introduction}
\begin{figure}[t]
\centering
\includegraphics[width=1.0\columnwidth]{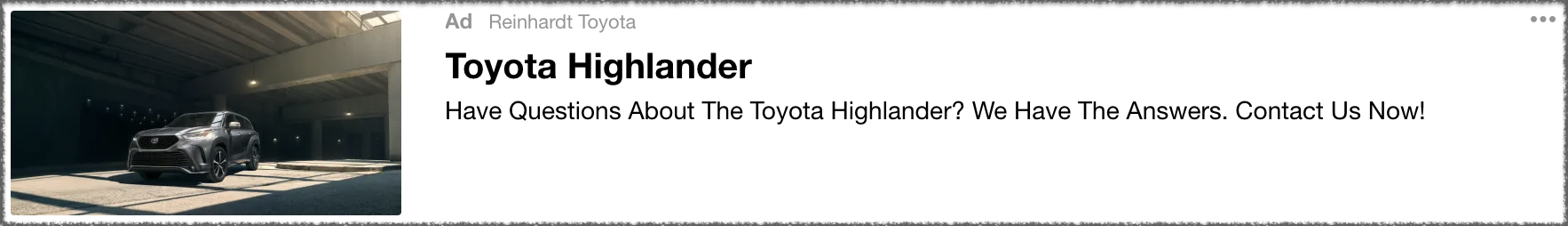}
\caption{A typical Yahoo Gemini native ad captured from Yahoo home-page stream. The ad consists of a title, an image, a description, and a transparent sponsorship notification.}
\label{fig: native ad}
\end{figure}

Yahoo Gemini native marketplace\footnote{See https://gemini.yahoo.com/advertiser/home} serves users with ads that are rendered to resemble the surrounding content (see Figure~\ref{fig: native ad} for a typical Gemini native ad). Operating with a yearly run-rate of several hundred million USDs\footnote{Due to commercial confidentiality matters, we report rough estimations or relative numbers regarding traffic volumes and revenue.}, Gemini native is one of Yahoo fastest growing and main businesses. With more than two billion impressions daily, and an inventory of a several hundred thousand active ads at any given time, this marketplace performs real-time \textit{first price}\alex{That's not up to date. We don't do GSP anymore}\oren{done} auctions that take into account budget considerations and targeting.

In order to rank the native ads for incoming users and their specific context according to the \textit{cost per click} (CPC) price type where advertisers pay for clicks, a score (or expected revenue) is calculated by multiplying the \textit{predicted click-through rate} (pCTR) by the bid for each active ad. The pCTR is provided by a model that is periodically updated by \offset\ - a feature-enhanced \textit{collaborative-filtering} (CF) based event-prediction algorithm %\cite{aharon2013off,aharon2017adaptive,aharon2019soft,arian2019feature}.
\cite{aharon2013off}\cite{aharon2017adaptive}\cite{kaplan2021dynamic}.
\offset is a one-pass\footnote{Goes through the data only once.} algorithm that updates its latent factor model for every new batch of logged data using a \textit{online gradient descent} (OGD) \alex{Online gradient descent (OGD)?}\oren{done} based learning approach.
\IEEEpubidadjcol 

\textit{Dynamic creative optimization} (DCO) is a recent product of the native portfolio Yahoo offers to its advertisers \cite{koren2020dynamic}. With DCO ads, advertisers may provide several assets for each native ad attribute, which are used to generate a plethora of combinations (e.g., having 3 titles, 2 images, and 3 descriptions, yields $3\times 2\times 3 = 18$ combinations or virtual native ads). Since different combinations may attract different crowds it may be beneficial to present certain combinations more frequently than others to maximize certain metrics such as \textit{click through-rate} (CTR) and revenue. The DCO product provides a simple and quick way for advertisers to identify the best ads that may be assembled from groups of assets without conducting a time consuming and expensive in-house A/B testings.

As advertisers spend is shifting toward \textit{optimize CPC} (oCPC) bidding strategy, where advertisers are interested in conversions\footnote{A conversion may follow (and be affiliated to) an impression (post-view) or a click (post-click), and may include events such as purchasing, registration, or app install.} but still pay for clicks, and since high CTR combinations do not necessarily provide high \textit{conversion rates} (CVR)\footnote{CVR is defined as the number of conversions divided by the number of impressions.}, we were required to provide a conversion based DCO solution that favors high CVR combinations. However, the event counting based approach of \cite{koren2020dynamic} that works well for clicks is problematic when conversions are considered, since CVRs are generally two orders of magnitude lower than CTRs.

In this work we present the approach used to serve the DCO best combinations for increasing CVR in Gemini native marketplace. To maintain similar system resources (i.e., model sizes and serving latency), we adhere to a two-stage solution (see \cite{aharon2019carousel}\cite{koren2020dynamic}). During the first stage, the Serving system (or Serving) conducts a regular auction where all combinations of a DCO ads participate as a single native ad \alex{where all combinations of a DCO ad participate as one item}\oren{done}. The second stage is invoked only in case a DCO ad wins the auction, and a DCO combination is drawn according to some combination distribution (a \textit{post-auction approach}). To calculate the combination distributions, we use an \textit{auxiliary combination CVR prediction model} which is not consumed by the Serving system and is used to provide the combination \textit{predicted CVRs} (pCVR) per certain traffic segments. The predictions are then turned into combination distributions where \alex{combinations having a higher pCVR}\oren{done} combinations having a higher pCVR are assigned with higher probabilities. As more conversions are accumulated, and certain DCO ad combinations are predicted to have higher CVRs, the distributions will drift from a uniform distribution, and the system will impress those combinations more frequently than others. The proposed approach may be seen as a simple and practical solution to the well know \textit{online decision problem} (see \cite[Chapter 2]{thompson2018now}), that is designed to satisfy both system requirements (such as model sizes and serving delay), and the inherent marketplace behaviour (such as time varying CVRs, and temporal trends).

After a short alpha testing phase, optimizing internal DCO ads, the system was launched for a beta testing phase, serving traffic of selected advertisers. The conversion based DCO solution was compared to a control bucket serving all combinations uniformly at random, and showed a staggering $53.5\%$ CVR lift. After the successful beta testing, the proposed conversion based DCO product is now fully deployed and available for all advertisers. The main contributions of this work are:
\begin{itemize}
\item We introduce a practical web scale optimization problem of maximizing native DCO ads CVR, under real-life constraints.
\item We present a post-auction DCO combination CVR prediction based approach. The proposed solution is simple, practical,
%and tailored to the existing Gemini native ad ranking system.
and easy to implement in any ad serving systems.
\alex{I think that the "tailored to" part is not a feature, but an anti-feature. I think we should emphasize that it's light weight, and easy to implement in ad serving systems}\oren{done}
\item Our solution was tested in scale, serving Gemini native users, and exhibits impressive CVR lifts in comparison to a control bucket serving all combinations uniformly at random.
\item The conversion based DCO system, is currently fully deployed, serving all Gemini native conversion ad traffic.
\end{itemize}

The rest of the paper is organized as follows. In Section~\ref{sec: background}, we bring relevant background, and cover related work in Section~\ref{sec:Related work}. We state the problem in Section~\ref{sec:problem statement}, and present our solution in Section \ref{sec:our approach}. Performance evaluation is considered in Section~\ref{sec:evaluation}. We conclude and discuss future work in Section~\ref{sec:Concluding remarks}.

\section{Relevant Background}\label{sec: background}
\subsection{{OFFSET} -- Event-Prediction Algorithm}\label{sec: offset}

Powering the Gemini native models is \offset (One-pass Factorization of Feature Sets): a feature enhanced collaborative-filtering (CF)-based ad event-prediction algorithm \cite{aharon2013off}\cite{aharon2017adaptive}\cite{aharon2019soft}\cite{arian2019feature}\cite{kaplan2021dynamic}. According to \offset, the \textit{predicted event probability} (pET) of a user $u$ and ad $a$ is roughly given by
\begin{equation}\label{eq:event probability prediction}
    \mathrm{pET}_{u,a} = \sigma(s_{u,a})\in [0,1]\ ,
\end{equation}
where $\sigma(x)=\left(1+e^{-x}\right)^{-1}$ is the \textit{Logistic sigmoid} function, and 
\begin{equation}\label{eq: score}
	s_{u,a}=b+\nu_{u}^T \nu_{a}\ ,
\end{equation}
$\nu_{u},\ \nu_{a}\in \R^D$ denote the user and ad \textit{latent factor} (LF) vectors respectively, and $b\in \R$ denotes the model bias. The product $\nu_{u}^T \nu_{a}$ indicates the tendency of user $u$ towards ad $a$, where a higher score translates into a higher pET. 
\Comment{It is noted that $\Theta=\{\nu_{u},\nu_{a},b\}$ are the model parameters which are learned from the logged data.} \alex{Well, they aren't. The feature latent vectors are the model's parameters}\oren{you are correct. I hope the reader is capable of this leap}.
It is noted that $\{\nu_{u},\nu_{a}\}$ are constructed from the model parameters $\Theta$ which are learned from the logged data and described in the sequel.

As mentioned above, \offset is used to drive many of Gemini native models, including (a) \textit{Click model} for predicting a click event (pCTR); (b) \textit{Conversion model} for predicting a conversion given a click event (pCONV); and (c) \textit{Ad close model} for predicting an ad close event (pCLOSE) \cite{silberstein2020ad}.

The ad and user LF vectors are constructed using their features to overcome data sparsity issues (ad events such as click, conversion, or close, are quite rare). For ads, a simple summation of their features LF vectors (e.g., ad id, campaign id, advertiser id, ad categories, etc.), all in dimension $D$, is used. The interaction between the different $d$-dimension user feature LF vectors to produce the user $D$-dimension LF vector is more complicated to support non-linear dependencies \alex{non-linear pairwise interactions?}\oren{I think it's good enough} between feature pairs. 

\begin{figure}[t]
\centering
\includegraphics[width=0.95\columnwidth]{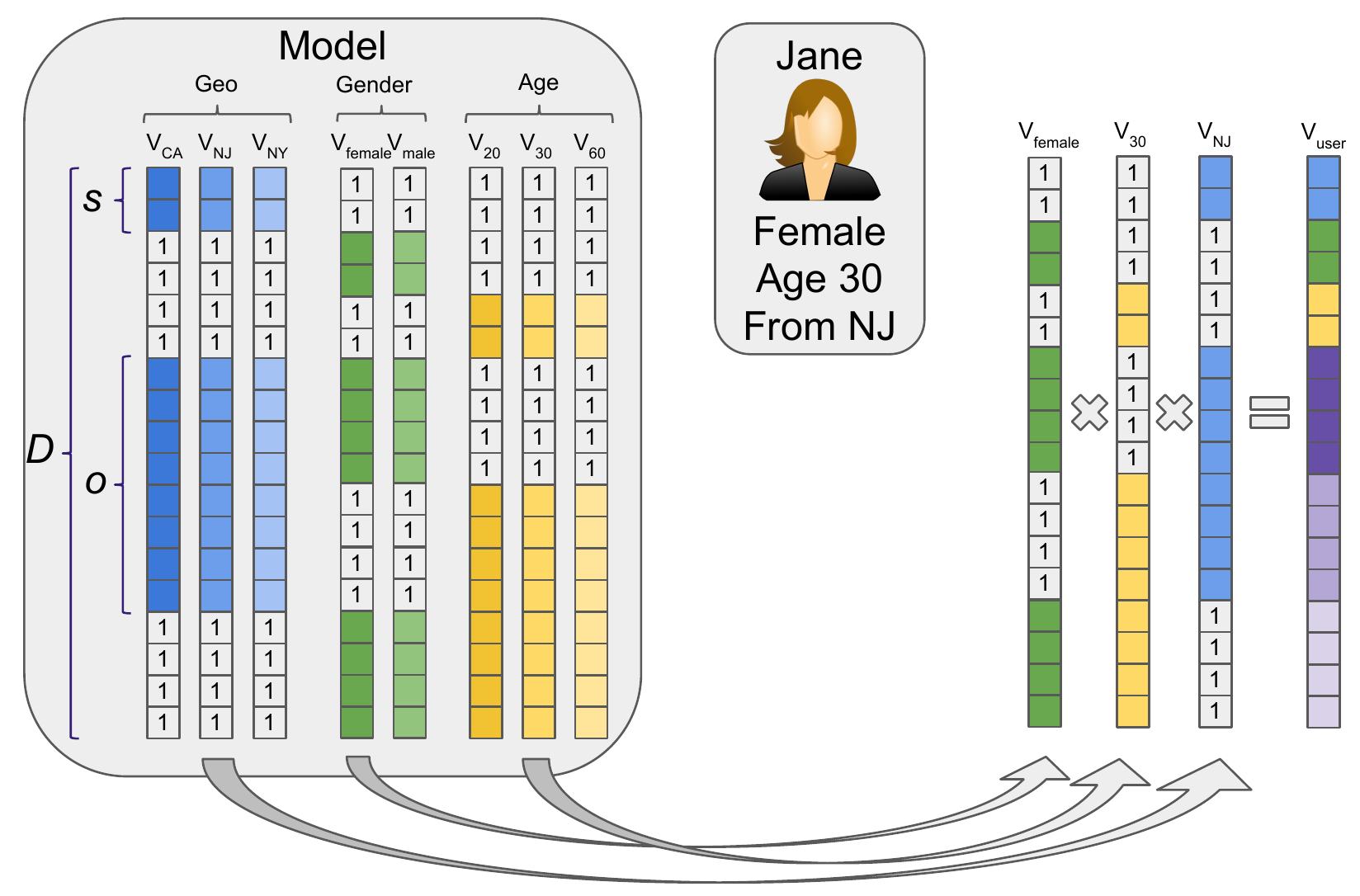}
\caption{Example of a user latent factor vector construction for $o=4,\ s=2$ and $K=3$ features (i.e., age, gender, and geo). The three relevant feature value vectors of dimension $d=10$ are retrieved from the model, appended with 1s, and multiplied entry wise to produce the final user vector of dimension $D=18$ on the right.}
\label{fig: user vector construction}
\vspace{0.1cm}
\end{figure}

The user vectors are constructed using their $K$-feature learned vectors $v_k\in \R^d$, $k\in\{1,...,K\}$ (e.g., gender values, age values, , device types, geo values, etc.). In particular, $o$ entries are allocated to each pair of user features, and $s$ entries are devoted to each feature vector alone. The dimension of a single feature value vector is therefore $d=(K-1)\cdot o + s$,
whereas the dimension of the combined user vector is $D=\binom{K}{2} \cdot o + K\cdot s$. An illustration of this construction is depicted in Fig.~\ref{fig: user vector construction}. The advantage over the conventional CF approach is that the model includes only $O(K)$ LF vectors, one for each feature value (e.g., three for gender - female, male, and unknown) instead of hundreds of millions of unique user LF vectors. Note that \offset may be seen as an enhancement of \alex{a special case of $\to$ an enhancement of}\oren{done} a \textit{factorization machine} (FM) \cite{rendle2010factorization} or as a variant of a \textit{field-aware factorization machine} (FFM) \cite{juan2017field}\cite{juan2016field}.

To learn the model parameters $\Theta$, \offset minimizes the \textit{logistic} loss (or LogLoss) of the training data set $\mathcal{T}$ (i.e., past negative and positive events) using a one-pass \textit{online gradient descent} (OGD)
\begin{equation*}
\argmin_{\Theta}\!\!\!\! \sum_{(u,a,y)\in \mathcal{T}} \mathcal{L}(u,a,y)\ ,
\end{equation*}
where $\mathcal{L}(u,a,y)$ equals
\Comment{\begin{multline}
\mathcal{L}(u,a,y)=\\-(1-y)\log\left(1-\mathrm{pEVENT}_{u,a}\right)-y \log \left(\mathrm{pEVENT}_{u,a}\right)+\frac{\lambda}{2}\sum_{\theta\in\Theta}\theta^2\ ,
\end{multline}}
\[
-(1-y)\log\left(1-\mathrm{pET}_{u,a}\right)-y \log \left(\mathrm{pET}_{u,a}\right)+\frac{\lambda}{2} \|\Theta\|_2^2\ ,
\]
\alex{In the formula above $\displaystyle \frac{\lambda}{2}\sum_{\theta\in\Theta}\theta^2$ $\to$ $\frac{\lambda}{2} \|\Theta\|_2^2$? More compact, and standard notation.}\oren{done}
$y \in \{0,1\}$ is the indicator (or label) for the event involving user $u$ and ad $a$, and $\lambda$ denotes the $L2$ regularization parameter. The OGD step sizes are determined by a variant of the \textit{adaptive gradient} (AdaGrad) algorithm \cite{duchi2011adaptive} (see \cite{aharon2017adaptive}).

The \offset algorithm applies an incremental training approach, where it continuously updates its model parameters with each batch of new training events (e.g., every 15 minutes for the click model, or 4 hours for the conversion model).
The \offset algorithm also includes an \textit{adaptive online hyper-parameter tuning mechanism} \cite{aharon2017adaptive}. This mechanism utilizes the parallel \textit{map-reduce} architecture of the Gemini native backend platform, and attempts to tune \offset hyper-parameters (e.g., OGD initial step size and AdaGrad parameters) to match the varying marketplace conditions (such as trend and temporal effects). We note that other components of \offset, such as similarity weights used for applying ``soft'' recency and frequency rules\footnote{How recent and how frequent a user may be impressed with the same ad or campaign.} are not described here for brevity (see \cite{aharon2019soft}).

\subsubsection{Weighted multi-value feature type}\label{sec:offset wmv feature}
The most generic feature type used by the \offset algorithm is the \textit{weighted multi-value feature} type. According to this feature type, the model includes a $d$-dimension LF vector for each of the $m$ feature values seen so far. In this case, the $d$-dimension vector of this weighted multi-value feature for user $u$ is\footnote{The definition is also valid for ad entities, however, the feature values LF vectors and resulting vector in this case are of dimension $D$.} 
\[
v=\frac{1}{\sqrt{n}}\sum_{i=1}^n w_{\ell_i}\ v_{\ell_i}\ ,
\]
\alex{Since we're not actually using the weights anywhere in this paper, it might be more beneficial to just describe multi-value features}\oren{I agree but I think it's ok}
where $\{\ell_i\}$ and $\{w_{\ell_i}\}$ are the $n$ feature values and accompanied weights associated with user $u$, and $\{v_{\ell_i}\}$ are the model LF vectors associated with values $\{\ell_i\}$. It is emphasized that the weights are given as part of the users' data and are not model parameters that are needed to be learned. Each time \offset encounters a new value for the feature, it assigns a random Gaussian vector to it with zero mean and covariance matrix $\eta\cdot I_d$, where $0\le \eta\ll 1$ and $I_d$ is the identity matrix\footnote{Other more sophisticated ``cold-start'' strategies are beyond the scope of this work.}. It is noted that simpler feature types such as \textit{categorical features} (e.g., age and gender), or non-weighted multi-value features (e.g., user category feature where all weights equal 1), are merely special cases of the weighted multi-value feature type.

\subsection{Serving ads}\label{sec:serving}
 When a user arrives at a Yahoo O\&O\footnote{\textit{Owned and Operated.} 
 site such as \textit{Yahoo news}, \textit{Yahoo mail}, and \textit{Yahoo finance}.} or Syndication\footnote{When Yahoo presents ads at third party site and shares revenue with the site owner.}\alex{Is the distinction between O\&O or Syndication important? I think it just adds redundant information which confuses the readers}\oren{It shows that we are a major exchange with our own sites as well as other sites} site, and a native slot should be populated by an ad, an auction takes place. At first, the Serving system generates a list of scores for all eligible active ads. In general, an ad's eligibility to a certain user in a certain context is determined by \textit{targeting}, which is beyond the scope of this work, and relates to user characterization (such as age, gender, and geo) specified by the advertiser to approach certain crowds.

\paragraph*{Auction}
The score is a measure that attempts to rank ads according to their expected revenue with respect to the arriving user and her context (i.e., her features, e.g., age, gender, geo, day, hour, site, device type, etc.). Roughly, an ad's score is defined as
\begin{equation}
\label{eq:rankingScore}
\mathrm{Score}_{u,a} = \mathrm{bid}_a \cdot \mathrm{pCTR}_{u,a}\ ,
\end{equation}
where $\mathrm{pCTR}_{u,a}$ (predicted click probability) is calculated by the \offset main click model (see Eq. \eqref{eq:event probability prediction} for a click event), and $\mathrm{bid}_a$
%\footnote{For simplicity, it is assumed that the bid depends on the ad only. However, it may be dependent on the users' context as well (e.g., the impression site).}
(in USD) is the amount of money the advertiser is willing to spend for a click on a \textit{manual cost per click} (mCPC) biding strategy ad $a$
%~\footnote{It is noted that also mCPC ads may have conversions.}. 
\alex{We'd like to stress that a DCO ad, along with all its assets, participates as one unit in the auction, and receives a single pCTR score.}\oren{I already added it before and it is mentioned explicitly in the DCO definition section}
For an \textit{optimized cost per click} (oCPC) biding strategy ad $\mathrm{bid}_a=\mathrm{pCONV}_{u,a}\cdot \mathrm{tCPA}$, where $\mathrm{pCONV}_{u,a}$ (predicted conversion probability given a click event) is calculated by the \offset main conversion model (see Eq. \eqref{eq:event probability prediction} for a conversion given click event), and the \textit{target cost per action} (tCPA) is the average cost the advertiser is expecting to spend on a conversion while still paying for clicks. Other, more sophisticated biding strategies such as \textit{enhanced optimized cost per click } (eCPC), where part of the traffic is used to estimate the conversion price, are outside of our scope.

According to the \textit{first price} auction, the winning ad is an eligible ad with the highest score $\mathrm{Score}_{u,a}$ (see \eqref{eq:rankingScore}). 

\Comment{To encourage advertiser truthfulness, the cost incurred by the winner of the auction is according to the \textit{generalized second price} (GSP) \cite{edelman2007internet}, which is defined as
\begin{equation}
\label{eq:gsp}
\mathrm{gsp} = \frac{\mathrm{Score}_{u,b}}{\mathrm{Score}_{u,a}} \cdot \mathrm{bid}_a = \frac{\mathrm{pCTR}_{u,b}}{\mathrm{pCTR}_{u,a}} \cdot \mathrm{bid}_b\ ,
\end{equation}
where $a$ and $b$ correspond to the winner of the auction and the runner-up, respectively. Note that by definition $\mathrm{gsp}\le\mathrm{bid}_a$, which means the winner will pay no more than its bid. Moreover, the winner will pay the minimal price required for winning the auction. In particular, if both ads have the same pCTR with respect to the user, the winner will pay the bid of the runner-up (i.e., $\mathrm{bid}_b$).}

\subsection{Native DCO ads}\label{sec:dco ads}
Native ads are designed to resemble the surrounding content of the site (or Web page) and are considered less intrusive than other ad types (see \cite{wojdynski2016native}). In general, Gemini native ads format usually includes several attributes such as title, image, and a short description (see Figure~\ref{fig: native ad} for a typical Gemini native ad). Other ad formats with richer structures such as Video and Carousel ads \cite{aharon2019carousel}, are outside the scope of this work.

\begin{figure}[t]
\centering
\includegraphics[width=0.95\columnwidth]{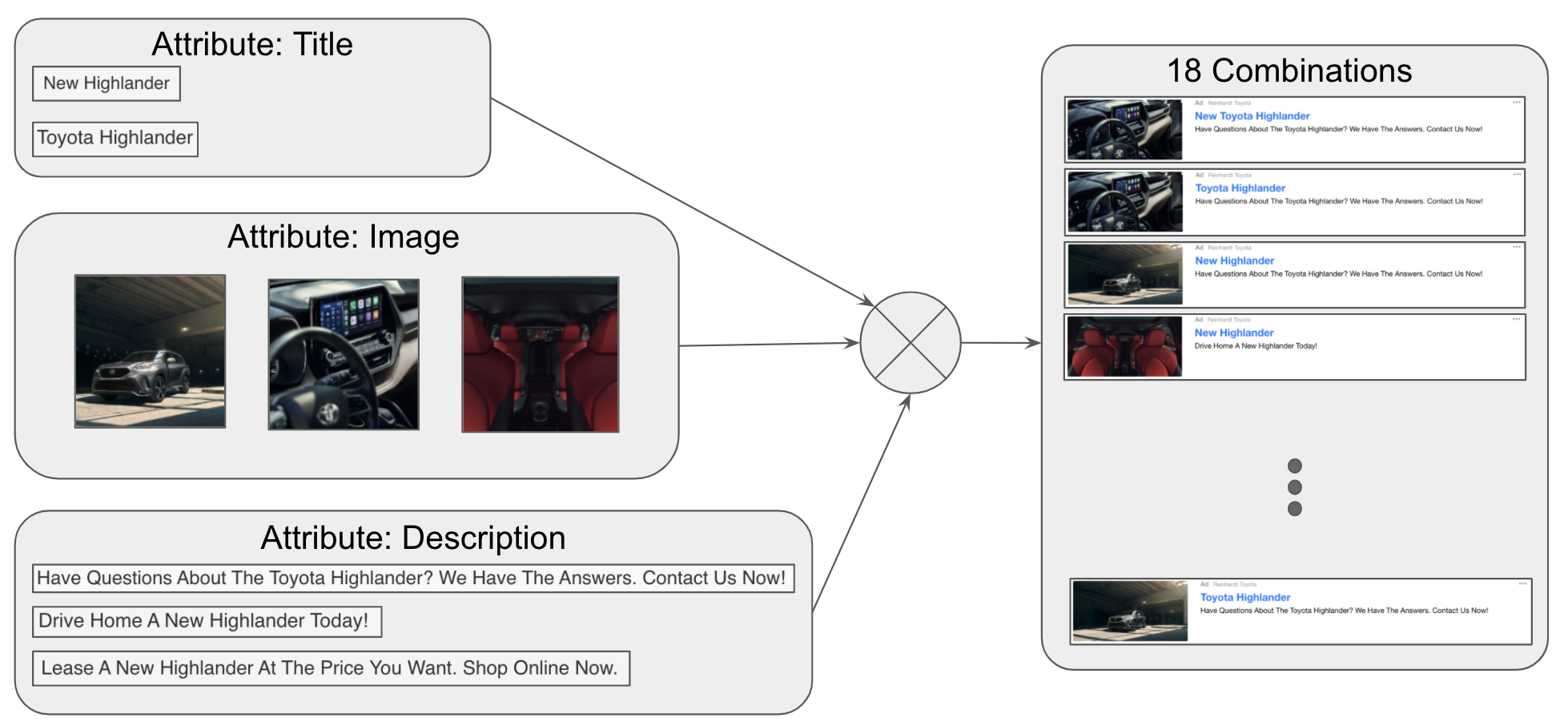}
\caption{A typical native DCO ad and its assets (2 titles, 3 images, and 3 descriptions) along with the resulting $18 = 2\times 3 \times 3$ combinations of virtual native ads.}
\label{fig: DCO ad}
\end{figure}

Two years ago, Yahoo started offering DCO native ads to its customers. With DCO ads, advertisers are allowed to upload up to 3 assets per each native ad attribute (e.g., 3 titles, 2 images, and 3 descriptions). This creates a plethora of combinations of ``conventional'' native ads that may be presented to users. In Figure~\ref{fig: DCO ad}, a real DCO ad and its assets are shown (i.e., 2 titles, 3 images, and 3 descriptions) along with the resulting $2\times 3\times 3 = 18$ combinations (or virtual native ads since combinations have no actual ad ID in the Gemini native system) that are generated by selecting one asset of each attribute. In case a DCO ad wins the auction and is about to be impressed, one of the combinations should be selected and rendered for presentation.

\section{Related work}\label{sec:Related work}
The problem of \textit{dynamic creative optimization} (DCO)\footnote{See https://en.wikipedia.org/wiki/Dynamic\_Creative\_Optimization .}, or finding the top combinations in terms of some metric such as CTR, CVR, or revenue, can be formulated as a reinforcement learning problem, where agents take actions to maximize some notion of cumulative reward \cite{sutton2018reinforcement}. In particular, our problem is closely related to the rich literature of the \textit{multi-armed bandit} (MAB) problem, which comes in several flavors, such as \textit{best arm identification} \cite{audibert2010best,karnin2013almost} and regret minimization \cite{bubeck2012regret}. The MAB problem is also studied in many settings, including mortal bandits \cite{chakrabarti2009mortal}, restless bandits \cite{slivkins2008adapting}, distributed bandits with feedback delay \cite{hillel2013distributed}, bandits with non-stationary rewards \cite{besbes2014stochastic}, and contextual bandits \cite{li2010contextual}. Our problem is also closely related to the \textit{online decision problem} \cite{thompson2018now}. Moreover, the use of an auxiliary prediction model to provoke actions is inspired by \textit{Thompson sampling}, which is widely used to address such problems (see \cite{chapelle2011empirical} for online advertising application). Other works that also adopt a holistic approach (as opposed to our post-auction approach) are \cite{hofmann2013balancing} and \cite{slivkins2013ranked}. Explore-exploit techniques are widely used to optimize Web content such as new items and ads. For example, \cite{agarwal2009online} describes an explore-exploit based system for selecting news items for Yahoo Homepage. There are also many commercial tools for conducting various ad related optimizations on social platforms (see \cite{lempel2012hierarchical} and references therein). It is also worth mentioning that similar DCO products are offered also by other online advertising providers such as Facebook, Google, and LinkedIn.

In the sequel we shall demonstrate that due to system requirements and constraints, our solution is based on a post-auction approach, which separates ad ranking and DCO ad rendering (as oppose to works that adopt a more holistic approach, e.g., \cite{agrawal2012analysis}\cite{hofmann2013balancing}\cite{li2010contextual}\cite{slivkins2013ranked}). The post-auction approach was used in \cite{koren2020dynamic} for a DCO product, however, with CTR as optimization goal. A \textit{successive elimination} algorithm based on robust CTR measurements was used to identify the ``best'' combinations (see also \cite{aharon2019carousel} for a similar setting which considers \textit{asset optimization} for Carousel ads). It is noted that although we consider a similar DCO product in the same native advertising marketplace as \cite{koren2020dynamic}, we are interested in maximizing CVR which is much lower than CTR in general. To overcome the extreme data sparsity condition, we apply a totally \textbf{different} \alex{radically different?}\oren{totally is not enough :-)} approach than that of \cite{koren2020dynamic}, which is based on \textit{event prediction} instead of \textit{event counting}. In addition, the proposed solution does not ``eliminate" combinations (as opposed to that of \cite{koren2020dynamic}) and inherently can follow combination CVR trends \alex{It's a contradiction. It cannot be stateless and follow trends. In fact, there is a state - the model which is used to generate the combinations}\oren{done}. Therefore, we believe the proposed CVR based DCO solution deserves its own separate reporting. 

Since CVR prediction is a major part of our solution, we mention a few works of the available vast literature of ad CVR prediction. An \offset based conversion-given-click prediction is described in \cite{aharon2015serving}. For more recent works on conversion rate prediction via post-click event modeling, see \cite{su2020attention}\cite{wen2020entire} and references therein.   

\section{Problem statement}\label{sec:problem statement}
Assume we have a DCO ad with $M$ attributes and $m_i$ assets for the $i$th attribute $i=1\ldots M$. Therefore, there are $N=\Pi_{i=1}^M m_i$ combinations (or virtual native ad). The goal is to use the additional \textit{degree-of-freedom} of having $N$ combinations, to maximize the DCO ad CVR and increase revenue.

It is noted that by maximizing the CVR (as opposed to CTR maximization \cite{koren2020dynamic}) we cannot guarantee a direct increase in revenue since according to the oCPC bidding strategy, advertisers still pay for clicks. However, since our main click and conversion-given-click prediction models are accurate, increasing the DCO ad CVR will cause the system to increase its bid (see oCPC explanation in Section \ref{sec:serving}) pushing DCO ads to win more auctions and spend more (if budget allows) \alex{It can also decrease bids, right? I'm not sure the statement is correct}\oren{if budget permits optimized bid will increase. The bottom line here is that immediate revenue lift is not guaranteed but long term lift is expected}. Moreover, for the mCPC bidding strategy DCO ads (see mCPC explanation in Section \ref{sec:serving}), the advertisers are expected to increase their budgets since they pay less for their conversions if CVR increases. Therefore, in the long-run, CVR improvement is expected to increase revenue. 

%Next, we discuss the main requirements, constraints, and setting conditions that dramatically affect the selected approach. 

\subsection{Requirements, constraints and conditions}\label{sec:requirements}
Since the conversion DCO system should be incorporated into the existing Gemini native architecture, there are several system requirements and inherent constraints that considerably affect the nature of the selected solution.  

Starting with the system requirements: (a) \textit{Model sizes} --
\offset click and conversion model sizes that are consumed by the Serving system cannot be considerably increased. Since most of the models consists of ad vectors, considering different combinations of DCO ads as new ads is not scalable; (b) \textit{Query-per-second (QPS) and service-level-agreement (SLA)} -- QPS cannot be considerably decreased (or alternatively SLA cannot be considerably increased). This means that the Serving system cannot make a query per DCO combination (e.g., using the combination number as an ad feature); (c) \textit{Serving complexity} -- computation complexity of the Serving system should be restricted to simple processing per ad (e.g., sigmoid function of the ads' score according to \eqref{eq:event probability prediction} and \eqref{eq: score}); and (d) for legacy reasons\footnote{Dictated by Gemini \textit{carousel asset optimization} \cite{aharon2019carousel}, and CTR based DCO \cite{koren2020dynamic} products.} the interface to the Serving system is limited to combination distributions per DCO ad and certain traffic segments (see Section \ref{sec: defining traffic segments}). The Serving system then uses the combination distributions in a predefined manner to draw the DCO combination for rendering the DCO ad. 

It is worth emphasizing that we are \textbf{extremely} sensitive to Serving system resources surge, as traffic is spread globally among many servers for maintaining harsh SLA requirements. However, we are almost \textbf{agnostic} to backend resources used to train our models, which is carried out over a couple of production grids.

Turning to the inherent constraints: (a) \textit{Delay} --
the time it takes for the system to update the model, index it, serve users, and log the users' actions\footnote{Applies to actions such as swipes and clicks but not to conversions.}. In the current production environment the average delay is more than an hour;
(b) \textit{Incremental mode} -- data processing is currently done in batches of a few minutes to a few hours worth of data (e.g., 15 minutes and 4 hours for click and conversion models, respectively). This means that approaches that require policy changes after each action cannot be considered. 

Finally, we describe the setting conditions that challenge the algorithm selection and system design: (a) \textit{Data sparsity} -- CVR is usually two orders of magnitude lower than CTR, hence, event counting based approach as the one used for optimizing click DCO ads \cite{koren2020dynamic} is problematic; (b) \textit{Reporting delays} -- conversion events may be reported up to 30 days after they occur. This is a real challenge that is hard to deal with since we do not want to wait so long before updating our models; (c) \textit{Varying impression rate} -- impression rates of DCO ads (as any other ad) are dictated mainly by their popularity, bid, and budget availability. These factors determine the ad delivery rate (the rate it wins auctions and gets impressions) and may change significantly over time; (d) \textit{Varying combination CVRs} -- CVR in general (as CTR) is known to vary over time (day over night, week days over weekends, etc.). In addition, popularity trends and ad fatigue may also cause ad (and combinations within DCO ads) CVR variations. The last two setting conditions make generating comparable combination CVRs measurements, a challenging task

%Combining the two aforementioned setting conditions makes generating comparable combination CVRs measurements, a challenging task.

\section{Our approach}\label{sec:our approach}
\subsection{Overview}
Due to the constraints dictated by the Gemini native architecture (see Section \ref{sec:requirements}) we chose to separate the processes of ad auction and DCO optimization, and adhere to a \textit{post-auction} approach. Hence, for each incoming impression we let the Serving system use \offset models to select the winning ad (see Section \ref{sec:serving}) and only then try to ``optimize'' the best combination for the current impression in case a DCO ad wins the auction. Since CVR is much lower than CTR, we use predictions instead of event counting to generate combination distributions. In particular, we train an \textit{auxiliary combination CVR prediction model} (see Section \ref{sec:auxiliary model}) and turn its predictions into combination distributions per DCO ad and certain \textit{traffic segments}, defined by a crossing contextual user features (see Section \ref{sec: defining traffic segments})\alex{, defined by a combination of user and contextual features, }\oren{done} at the end of every training period. Then, the distributions (packed in a DCO model file) are periodically sent to the Serving system, allowing it to draw the selected combination according to the relevant distribution before rendering the winning DCO ad. It is noted that we maximize CVR (probability of a conversion given an impression event) and not the probability of a conversion given a click event. We do that since after the auction, when the combination distributions are used by Serving, the impression has not happened yet and therefore CVR is the correct term to maximize. \alex{We'd like to emphasize that the novelty of our approach stems not from a novel conversion prediction model, but from our utilization of an existing well-proven conversion prediction framework to facilitate our light-weight post-auction approach to overcome extreme data sparsity issues (this is to address a reviewer's comment)}\oren{done}

We would like to emphasize that the novelty of our approach stems not from a novel conversion prediction model, but from our utilization of an existing well-proven conversion prediction framework to facilitate our light-weight post-auction approach to overcome challenges such as extreme data sparsity condition (see Section \ref{sec:requirements}).

\subsection{Traffic segments definition}\label{sec: defining traffic segments}
We define the traffic segments by configurable segment keys that relate to categorical unweighted user features
%\footnote{For simplicity we use only categorical unweighted features.} 
(e.g., age, gender, and device). Assuming we use gender (i.e., male, female, and unknown) and device (i.e., mobile, desktop, tablet, and unknown) as segment keys, then for each DCO ad we have $3\times 4=12$ segments (e.g., female $\times$ desktop). 
For each DCO logged impression and conversion events, the segment keys are extracted and used as user features for training the auxiliary prediction model (see Section \ref{sec:auxiliary model}). During serving time, in case a DCO ad wins the auction, the segment keys are extracted from the incoming user's features, and used to locate the corresponding traffic segment distribution for drawing the combination and rendering the DCO ad (see Section \ref{sec:rendering DCO ads in serving time}).

\subsection{Rendering DCO ads during serving time}\label{sec:rendering DCO ads in serving time}
Assume the auction winning ad is some eligible DCO ad $A$ with $N$ combinations $\mathcal{C}_A=\{C_n\}_{n=1}^N$. The Serving system extracts the segment keys of the incoming user's features to determine the traffic segment $S$ and locates the corresponding combination distribution $Q_{A,S}=\{Q_{C_n}\}_{n=1}^N$ in the DCO model file. Then, it draws one combination according to $Q_{A,S}$ before rendering $A$.

\subsection{\hspace{-0.1cm}Auxiliary combination CVR prediction model}\label{sec:auxiliary model}
To predict DCO ads combinations' CVRs, we train an \offset based event prediction model (see Section \ref{sec: offset}). \alex{Consider, instead of the prev. sentence: "To generate the combination distributions, we train an \offset based model model for predicting pCVR, and transform its predictions to distributions."}\oren{something in the middle}.
\begin{itemize}
\item Impressions are used as negative events while conversions (both post-click and post-view) are used as positive events.
\item Since conversions are reported with long delays (up to 30 days after the actual view or click), we do not join the conversion and the impression that entailed it before training the model. Hence, each positive event is also trained as a negative event. Therefore, CVRs are slightly under-predicted and their predictions should be bias-corrected before turned into actual distributions (see Section \ref{sec: prediction correction}).
\item To reduce the backend resources required to train the auxiliary prediction model we downsample the negative events (i.e., impressions) before training. Hence, predictions should be bias-corrected \alex{consider ``corrected'' $\to$ bias-corrected}\oren{done} also to account for this downsampling operation \alex{consider: for this downsampling operation $\to$ to account for this downsampling}\oren{done} before turned into actual distributions (see Section \ref{sec: prediction correction}).
\end{itemize}
It is noted that unlike the main click and conversion-given-click models that are consumed (or indexed) by the Serving system for ranking the ads (see Section \ref{sec:serving}), the auxiliary model is used only for generating DCO combination distributions. In addition, we train the auxiliary model over \textit{all} conversion ad traffic and not only DCO ad traffic for triggering collaborative filtering patterns that help in ``filling'' the gaps due to conversion data sparsity.

Next, we consider the ad and user features used to train the auxiliary model, and tie the latter to the traffic segments defined in Section \ref{sec: defining traffic segments}.

\subsubsection{User features}\label{sec:user features}
Since the auxiliary model is used to predict the DCO ads combinations' CVRs for each traffic segment of interest (see Section \ref{sec: defining traffic segments}), we use the corresponding segment keys as user features. For example, if segment traffics are defined by \textit{gender} and \textit{device}, then we use those as user features. 

It is noted that adding more user features may provide more accurate CVR predictions. However, there is little benefit in doing that in our setting, since we are interested in the CVR predictions of certain traffic segments that are defined by certain segment keys. Therefore, even if we add more features, we would have to average the predictions over the extra features' values to get the predictions over the traffic segments of interest, and actually ``lose'' the improved accuracy. It is concluded that in our setting it is sufficient to set the traffic segment keys as user features.     

\subsubsection{Ad features}\label{sec:ad features}
We may use all ``standard'' ad features available for Gemini native models, such as \textit{ad ID\ \footnote{ID is a unique string that is generated by the system and assigned to every ad related entity, such as the ad itself, its assets, its campaign, and its advertiser.}}, \textit{campaign ID},  \textit{advertiser ID}, and \textit{ad category} (a multi-value feature with few dozens of values such as \textit{sports} and \textit{electronics} - see Section \ref{sec:offset wmv feature}). To predict the CVR for each combination of each DCO ad, we require a special ad feature that ties together the assets that form each combination.

\paragraph{Combination assets ad feature}\label{sec:assets feature}
Each event (impression or conversion) which involves a DCO ad, includes information regarding the actual assets used to render the ad. For example, assuming a certain DCO ad was impressed using a certain combination with description ID - De123, image ID - Im456, and title ID - Ti789, then a weighed multi-value feature \{(De123,1),(Im456,1),(Ti789,1)\} with unit weights (see Section \ref{sec:offset wmv feature}) is generated and used to train the auxiliary model. It is noted that assets multi-value representation is used instead of representing the combination number as an ad feature, to create dependencies among the DCO ad's combinations that share assets (e.g., share the same title and description but include different images). This in turn triggers collaborative patterns that help ``filling'' the gaps caused by data sparsity. Finally, for all events that involve non-DCO ads, the feature is assigned with ``NONDCO'' value and a unit weight, i.e.,  \{(NONDCO,1)\}. 

\subsection{Turning predictions into distributions}\label{sec: predictions to distributions}
At the end of every training period of the auxiliary prediction model (e.g., 4 hours), we scan it for DCO ads. Then, for each DCO ad we use the auxiliary model to calculate the DCO ad combinations' CVR predictions for each traffic segment. To do that we gather the particular DCO ad features values (see Section \ref{sec:ad features}), query the model, and construct all the DCO ad's combinations LF vectors (see Section \ref{sec: offset}). Since the system is designed such that traffic key segments coincides with the auxiliary prediction model user features, to get the user LF vector we query the model using the segment keys values that define the segment (e.g., female and desktop for gender and device segment keys), and construct the user vector (see Section \ref{sec: offset}). After we have the user and combinations LF vectors, we extract the model bias and use \eqref{eq:event probability prediction} to calculate the combinations' CVR predictions. Next, the predictions are bias-corrected to compensate for the \textit{non-join} and impression \textit{downsampling} operations (see Section \ref{sec: prediction correction}). Finally, we use \textit{SoftMax}\footnote{See https://en.wikipedia.org/wiki/Softmax\_function .} to translate the true predictions into distributions and add a uniform component to produce the final distribution. All distributions are gathered into a DCO distributions file (or table) and sent to the Serving system to be used during serving time (see Section \ref{sec:rendering DCO ads in serving time}).

See Algorithm \ref{algo:P2D} for a formal description of the \textit{predictions to distributions} (P2D) algorithm.
%Other actions such as \textit{handling ads with fewer conversions}, \textit{handling traffic segments with low maximal predictions}, and \textit{omitting inactive DCO ads} are omitted from Algorithm~\ref{algo:P2D} for clarity.
\begin{algorithm}%[htb]
\caption{Predictions to distributions (P2D)} \label{algo:P2D}
\textbf{Input:}\\
$\Theta$ - latest auxiliary DCO combination CVR prediction model\\
$\lambda\in[0,1]$ - uniform component parameter\\
$\beta>0$ - \textit{SoftMax} parameter\\
$r_{ds}$ - impression downsampling factor\\ 
\textbf{Output (updated after each training period):}\\
$T_{DCO}$ - DCO combination distributions table\\ %. A list of triplets (ad, traffic segment, distribution)\\
\vspace{-0.3cm}
\begin{algorithmic}[1]
\STATE query $\Theta$ for all DCO ads $\mathcal{A}$, and model bias $b$
%\STATE \quad get the bias $b$ 
%\STATE \quad get all DCO ads $\mathcal{A}$ 
\STATE generate a Cartesian product set of all values of all user features $\mathcal{F}= F_1\times\ldots\times F_n$
\STATE generate all user LF vectors (one per traffic segment)\\ $\nu_u\ ,\ u\in \mathcal{F}$
%\vspace{-0.1cm}
\FOR{each DCO ad $A\in\mathcal{A}$}
\STATE query $\Theta$ 
\STATE \quad extract all ``standard'' ad $A$ features' values $g_1,\ldots,g_{m-1}$
\STATE \quad generate a Cartesian product set of all assets of ad $A$ attributes $\mathcal{C}_A = Att_1\times\ldots\times Att_k$ (all combinations)
\FOR{each user LF vector (traffic segment) $\nu_u\ ,\ u\in\mathcal{F}$}
\FOR{each combination $C\in \mathcal{C}_A$}
\STATE use $C$ to generate the respective \textit{combination assets ad feature} $g_C$
\STATE query $\Theta$ and get the ad combination LF vector\\ $\nu_a\ ,\ a=\{g_1,\ldots,g_{m-1},g_C\}$
\STATE calculate the combination CVR prediction\\ $P'_C=\sigma(b+\nu_u\ \nu_a^T)$
\STATE calculate the ``true'' prediction\\ $P_C=\max{\left\{1,\frac{P'_C}{r_{ds}(1-P'_C)}\right\}}$
\ENDFOR
\STATE find the maximal prediction $P_M=\max{\{P_C\}}$
\Comment{\IF{$P_M<P_{min}$}
\STATE \bf{CONTINUE}
\ENDIF}
\STATE turn predictions $\{P_C\}$ to distribution using \textit{SoftMax} function and a uniform component\\
\vspace{-0.3cm}
\[
Q_C = (1-\lambda)\frac{e^{-\beta (1-P_C/P_M)}}{\sum_{S\in \mathcal{C}_A}e^{-\beta (1-P_S/P_M)}}+\frac{\lambda}{\abs{\mathcal{C}_A}}\ ,\ \forall C\in \mathcal{C}_A
\] \alex{I think you forgot to change $(1 - \lambda)$ to $\left(1 - \frac{\lambda}{|\mathcal{C}_A|}\right)$}\oren{not forgotten. If you summarize all probabilities you get 1}
\vspace{-0.2cm}
\STATE add the triplet $\{A,u,\{Q_C\}\}$ to $T_{DCO}$
\ENDFOR
\ENDFOR
\end{algorithmic}
\end{algorithm}

\Comment{The following actions are omitted from Algorithm~\ref{algo:P2D} for clarity and are mentioned here instead.
\begin{itemize}
    \item \textit{Handling ads with fewer conversions} - DCO ads with less than a predefined number of conversions $N_c$ (a system parameter, e.g., $N_c=1$) are assigned with uniform distributions for all traffic segments.
    \item \textit{Handling traffic segments with low maximal prediction $P_M$} - traffic segments with $P_M<P_{min}$ (a system parameter, e.g., $P_{min} = 1e-9$) are assigned with uniform distribution.
    \item \textit{Omitting inactive DCO ads} - to reduce the auxiliary prediction model size and training time, we omit ads (DCO and non-DCO) that were inactive (i.e., had neither impressions nor conversions) for a predefined time period (e.g., a week) and treat them as new ads if they reappear in traffic.
\end{itemize}}

\subsubsection{Why \textit{SoftMax}?}\label{sec:why softMax}
Our choice of using \textit{SoftMax} facilitates a controlled mechanism to provide an explore-exploit trade-off and lets the system follow trends while presenting ``better'' combinations more frequently. However, there are other solutions that hold similar characteristics, e.g., we could use the predictions and select the closest distribution on the unit simplex, or just normalize each prediction by the sum of predictions. In the sequel we provide a theoretical justification to our choice.

Assuming a conversion DCO ad $A$ has won the auction for an impression that belongs to a traffic segment $u$, we would like to maximize the chance for getting a conversion by exploiting the additional degree-of-freedom and select the combination that has the best chance to entail a conversion. Accordingly, we would like to maximize
\vspace{-0.18cm}
\begin{multline}\label{eq:optimization}
Pr(\mathrm{conversion}|u,A)=\\ \sum_{C\in \mathcal{C}_A}\ Pr(\mathrm{conversion}|C\  \mathrm{selected},u,A)Pr(C\  \mathrm{selected}|u,A)\ ,
\end{multline}
where the probability $Pr(\mathrm{conversion}|C\ \mathrm{selected},u,A)$ can be approximated by the auxiliary model true CVR prediction $P_C$ (see step 13 of Algorithm \ref{algo:P2D}), and the probability $\Pr(C\ \mathrm{selected}|u,A)$ is the traffic segment combination probability $Q_C$ we need to set for DCO ad $A$. A trivial solution to maximize \eqref{eq:optimization} is to assign all the probability mass to the combination that has the highest CVR prediction. However, presenting the same combination repeatedly to all users falling into the traffic segment (every time DCO ad $A$ wins the auctions) is undesired. This is since it might enhance the ad fatigue phenomenon and also prevent the system from exploring and following trends causing other combinations to be attractive in terms of CVR. Therefore, we would like to ``pull'' the distribution towards a uniform distribution in a controlled manner and create a trade-off between exploration and exploitation. A natural way to do the latter is to add an \textit{Entropy} regularization term into the optimization problem of \eqref{eq:optimization} and rewrite it as
\vspace{-0.2cm}
\begin{multline}\label{eq:optimization with regularization}
Q^* = \argmax_Q \left\{ \sum_{C\in\mathcal{C}_A}P_C\ Q_C - \alpha \sum_{C\in\mathcal{C}_A} Q_C \ln(Q_C)\ :\ \right.\\ \left. \sum_{C\in\mathcal{C}_A} Q_C = 1,\ Q_C\geq 0\ \forall{C\in\mathcal{C}_A}, \right\}\ ,
\end{multline}
where the two constraints are added to ensure $Q=\{Q_C\}$ is a distribution function (in a vector representation), and $\alpha>0$ is the regularization parameter. Next, we apply a well known result in convex analysis \cite[Section 4.4.10]{doi:10.1137/1.9781611974997} that the \textit{convex conjugate} of $\phi(z) = \sum_{i=1}^d z_i \ln(z_i)$ defined over the \textit{unit simplex} is\footnote{Note that the result of \eqref{eq:convex conjugate} is a scalar while the result of \eqref{eq:optimization with regularization} is a vector.}
\begin{equation}\label{eq:convex conjugate}
\phi^*(x) = \max_z \left\{ x^T z - \phi(z) : \sum_{i=1}^d z_i = 1,\ z \geq 0 \right\} = \ln \left( \sum_{i=1}^d e^{x_i} \right)\ ,
\end{equation}
and according to the \textit{conjugate-subgradient} theorem \cite[Theorem 4.20]{doi:10.1137/1.9781611974997}, we have
\[
Q^* = \left.\left(\nabla \phi^*(x)\right)\right |_{x=P/\alpha} = \frac{\{e^{P_C/\alpha}\}}{\sum_{C\in\mathcal{C}_A}e^{P_C/\alpha}} = SoftMax(P/\alpha)\ .
\] 
where $P=\{P_C\}$ are the combinations' predictions (in a vector presentation). Setting $\alpha=P_M/\beta$ \alex{this is something I don't understand - I know it works in practice, but why actually the entropy regularization parameter $\alpha$ should depend on the maximum $P_C$?}\oren{it's mathematically correct and meant to support an explainable relative presentation of the assigned probabilities, i.e., the numerator is 1 for $P_M$ and $<1$ for all the rest} where $P_M=\max \{P_C\}$, and recalling that \textit{SoftMax} is invariant under shift, so
\[
SoftMax\left(\beta P/P_M\right)=SoftMax\left(-\beta\left(1-P/P_M\right)\right)\ ,
\]
we get the \textit{SoftMax} combination distribution component of  Algorithm \ref{algo:P2D} (step 16). The final presentation of the \textit{SoftMax} argument is preferred to allow relative interpretation, e.g., setting $\beta=6.93$ means that a $10\%$ difference between the ``best'' combination and its runner-up entails approximately twice the probability. It is noted that the uniform component $\lambda$ is added to ensure a minimal amount of exploration. This constraint may be also added to the optimization problem, however, for clarity matters it is added in a straightforward sub-optimal manner.

\subsubsection{Prediction correction}\label{sec: prediction correction}
As mentioned earlier we downsample the impressions (e.g., factor $r_{ds}=100$) to reduce the auxiliary model training resources. In addition, we also do not ``join'' conversions with their impressions to avoid long training delays (conversions may be reported up to 30 days after they occur). Therefore, to get the ``correct'' CVR prediction we have to compensate for these two actions. The following is a rough approximation that is accurate only on average. Assuming we have $V$ conversions and $S$ skips (i.e., impressions without conversions) for a certain ad. Then, the average ``raw'' CVR with \textit{downsampling} of $r_{ds}$ and \textit{non-join} operation may be written as
\[
\mathrm{CVR'}=\frac{V}{V+\frac{V+S}{r_{ds}}}\ .
\]
Since the correct average $CVR=V/(V+S)$,
%\[
%\mathrm{CVR}=\frac{V}{V+S}\ ,
%\]
it is easily verified that
\begin{equation}
    \mathrm{CVR}=\min\left\{1,\frac{\mathrm{CVR'}}{r_{ds}(1-\mathrm{CVR'})}\right\}\ ,
\end{equation}
where the minimum operation is required to keep $\mathrm{CVR}<1$ for very high average ``raw'' conversion rates $CVR'>r_{ds}/(1+r_{ds})$ which is $\sim 0.99$ for $r_{ds}=100$.
%\[
%\mathrm{CVR'}>\left.\frac{r_{ds}}{1+r_{ds}}\right|_{r_{ds}=10%0}\approx0.99\ ,
%\]

\subsection{Benefits of our approach}\label{sec: benefits of our approach}
\begin{itemize}
    \item The post-auction approach is selected for its reduced system complexity; As it does not increase the main click and conversion model sizes (consumed periodically by the Serving system and used in serving time), nor does it reduce the serving QPS (see Section~\ref{sec:requirements}). This is since the combination selection is done after the auction and no further index queries are required to rank the combinations. Instead, only a simple draw based selection, done according to the combination distribution, is applied.
    \item Unlike the CTR based DCO solution (reported in \cite{koren2020dynamic}), which uses event (i.e., clicks and skips) counting, the current solution uses combination CVR prediction provided by an auxiliary CF model capable of revealing collaborative conversion patterns that help in ``filling'' the gaps caused by data sparsity. 
    \item The solution is able to follow and discover trends that affect the attractiveness of different combinations throughout the ad's lifespan. In addition, it can combat the ad fatigue effect by reducing the probabilities of combinations the users are weary of, and increasing the probabilities of other less exploited combinations.
    \item Unlike the CTR based DCO successive elimination solution (reported in \cite{koren2020dynamic}), the current solution does not eliminates combinations and therefore may be easily trained on traffic of other candidate models. This makes the ramp-up process of such a model rather easy in comparison to that of the CTR based model of \cite{koren2020dynamic}. 
    %\oren{no need for assumptions like constant combination ranking}.
\end{itemize}
We do not claim that our solution is optimal, however, as we shell demonstrate in the sequel, we do manage to utilize the additional degree-of-freedom and considerably increase the DCO ads' CVRs.

\Comment{
\subsection{Auxiliary prediction model offline evaluation}
\subsubsection{Settings}
\subsubsection{Metric}
\subsubsection{Baselines}
\subsubsection{Results}
}

\section{Performance Evaluation}\label{sec:evaluation}
In this section we report the online performance of the conversion based DCO system. We describe the setting, define the performance metrics and baselines, present the results and discuss them.

It is noted that since proprietary data is used for evaluating our solution, it is evident that others can not reproduce our results. This caveat is common in papers describing commercial systems and should not undermine the overall contribution of the work. Moreover, due to commercial confidentially matters we report performance lifts only (e.g., CVR lift). However, this should not affect the trends and observations we present.

\paragraph{\textbf{Settings}}
After a phase of offline evaluation used to select the auxiliary prediction model features and parameters, the conversion based DCO system was launched for an alpha online test phase, optimizing internal ads. The online tests demonstrated the potential of the proposed solution and were also used for tuning various system parameters. As a result, we trained the auxiliary model with all available ad features (see Section \ref{sec:ad features}). As user features we use \textit{gender} and \textit{device} that are also used as traffic segment keys (a total of $3\times 4=12$ segments for each DCO ad). In addition, we set the user features independent and overlapping LF vector sizes to $s=o=12$, so the final ad and user LF vector sizes are $D=36$ (see Section \ref{sec: offset}). It is noted that other auxiliary model training parameters such as OGD step-size, AdaGrad parameters, and regularization parameter, are tuned automatically by \offset built-in online hyper parameter tuning mechanism (see Section \ref{sec: offset} and \cite{aharon2017adaptive}). For the P2D algorithm (see Algorithm \ref{algo:P2D}), we set the \textit{SoftMax} factor $\beta=13.86$ (so a $10$\% predicted CVR difference between the ``best'' and runner-up combinations results in a probability ratio of $4$), and the uniform component $\lambda$ times the number of combinations to $0.1$ (i.e., 10\% of the probability mass). The aforementioned parameter set demonstrated the best CVR lifts we experienced in the limited grid search we performed during our online alpha test phase.

After demonstrating good online performance, the system was pushed to production (serving $90\%$ of all traffic) for a beta test phase, done with a few selected advertisers, and included several dozens of conversion DCO ads with a total daily average of over million impressions. Although the beta experiments were coordinated, the advertisers controlled all aspects of their conversion DCO ads such as ad content, daily budget, and target CPA (tCPA).

\Comment{
\paragraph{\textbf{Baseline}}
At any stage we compared the conversion based DCO bucket metrics to those of a \textbf{control bucket} operating with no conversion based DCO system, and rendering DCO ads using combinations that are chosen uniformly at random. The uniform DCO bucket was deployed on the remaining $10\%$ of the traffic. We note that since we deal with a real Web scale system and not an offline experimental system, implementing and deploying other more sophisticated baselines is problematic. Lastly, it is worth mentioning that using our CTR based DCO system of \cite{koren2020dynamic} as baseline is not informative since the two systems optimize different metrics.}

\paragraph{\textbf{Baselines}}
At any stage we compared the conversion based DCO bucket metrics to those of two \textbf{control buckets} each serving $5\%$ of the traffic. The \textbf{Uniform bucket} was operating with no conversion based DCO system, and rendering DCO ads using combinations that are chosen uniformly at random. The \textbf{CTR bucket} was operating the CTR based DCO system of \cite{koren2020dynamic} which is aimed at maximizing the DCO combinations' CTR. 
We note that since we deal with a real Web scale system and not an offline experimental system, implementing and deploying other more sophisticated baselines is problematic.

\Comment{
\begin{table}%[htb]
%\scriptsize
\begin{center}
    \begin{tabular}{|c|c|c|c|c|c|}
    \hline
    Test period & CVR lift & CTR lift & Delivery lift & CPM lift & CPA lift\\
    \hline\hline
    28 days & 53.5\% & 1.69\% & 0.01\% & 0.44\%  & -34.6\% \\
    \hline
    \end{tabular} 
    \end{center}
    \caption{Conversion based DCO test phase results summary.}\label{table: bucket results}
\end{table}}

\begin{table}%[htb]
%\scriptsize
\begin{center}
    \begin{tabular}{|c|c|c|c|c|c|}
    \hline
    baseline & CVR lift & CTR lift & Delivery lift & CPM lift & CPA lift\\
    \hline\hline
    Uniform & 53.5\% & 1.69\% & 0.01\% & 0.44\%  & -34.6\% \\
    \hline
    CTR &  38.0\% & -0.92\% & 0.00\% & -3.57\%  & -30.1\% \\
    \hline
    \end{tabular} 
    \end{center}
    \caption{Conversion based DCO test phase results summary.}\label{table: bucket results}
    \vspace{-0.5cm}
\end{table}

\paragraph{\textbf{Metrics}}
To evaluate our system we use the following. 
\begin{itemize}
    \item CVR - conversion rate, i.e., the number of conversions divided by the number of impressions, where higher is better. Increasing the CVR metric is the goal of the system. We expect positive lifts\footnote{Lifts are calculated by $(M_{DCO}/M_{baseline}-1)\cdot 100$, for any positive metric $M>0$.} if our system works properly.
    \item CTR - click through rate, i.e., the number of clicks divided by the number of impressions, where higher is better. We do not expect a major CTR lift over the Uniform baseline, since our system is not designed to maximize CTR. Moreover, we do expect a CTR drop in comparison to the CTR baseline which is designed to maximize CTR.
    \item Delivery - the number of impressions of the targeted traffic (i.e., DCO traffic). The A/B testing mechanism ensures that different buckets will get their fair share of incoming impressions opportunities. However, the amount of DCO traffic falling into each bucket depends on the results of the auctions which happens only after the incoming impressions were distributed to the different buckets.
    \item CPM - cost per thousand impressions, i.e., the total cost divided by the number of impressions and multiplied by 1000, where higher is better.
    \item CPA  - cost per action, i.e., the total cost divide by the number of conversions, where lower is better. Lower CPA means advertisers spend less for each conversion\alex{on each conversion}\oren{done}.
\end{itemize}
While the CVR, CTR, and Delivery metrics, are easily measured and interpreted, the CPM and CPA metrics, as we shall explain in the sequel, should be handled more carefully.

\paragraph{\textbf{Results and discussion}}\label{sec:results}
The results of the conversion based DCO system beta test are summarised in Table \ref{table: bucket results}. We emphasize that the reported results were measured over the traffic of all DCO ads (oCPC and mCPC) that had at-least one conversion during the evaluation period, while other DCO ads traffic is ignored. Over a period of 28 days (four weeks), we measured an impressive $53.5\%$ and $38.0\%$ average CVR lifts of the conversion based DCO bucket in comparison to the Uniform and CTR buckets, respectively. Since the performance were measured over millions of impressions and the lift is large, the superiority of our approach over the baselines is established with high confidence. In particular, the positive CVR lifts are statistically significant with $p-value < 10^{-14}$. \alex{I think that since the statistical model doesn't take delay into account, i.e. you aren't modeling the hazard function, and just using the #conv / #impr, I think this number isn't very informative. If I were a reviewer, it would negatively impress me. But I don't have that much experience, so it's up to you.}\oren{you are probably correct but the reviewers probably won't notice and p value report is expected}

In addition to the CVR hefty lift we also measured a $1.69\%$ average CTR lift ($p-value=0.016$) over the Uniform baseline and only a $-0.92\%$ CTR drop (negative lift) when compared to the CTR baseline, which means that there is some positive correlation between DCO combinations that entail more conversions and those that entail more clicks. However, it is quite evident that the system is indeed tuned for optimizing CVR. Moreover, the measured CTR drop in comparison to the CTR baseline is expected since the latter is designed to maximize CTR.

Examining the minuscule Delivery lifts of $0.01\%$ and $0.00\%$, it is concluded that the targeted traffic of DCO ads with at least one conversion is almost equal (after proper normalization) between the conversion based DCO and baseline buckets. This is easily explain by recalling that both buckets rank the ads using the same main click and conversion-given-click prediction models which are trained over the entire traffic. Moreover, the DCO combination selection is done after the auction and has no direct influence on the delivery. It is also noted that since normalized delivery is almost equal between the all buckets, CPM lift which is considered next, is actually revenue lift.

The measured CPM lift of $0.44\%$ over the Uniform bucket seems quite disappointing and does not reflects the staggering CVR lift. However, the fact that all buckets use the same main models for ad ranking and therefore have similar deliveries, combined with the modest CTR lift and the fact that all DCO ads pay for clicks, explains the CPM result. Moreover, the baselines actually benefit from the increased CVR of DCO ads caused by the conversion based DCO system. This is since the main ranking models are trained over the entire traffic which is dominated by the conversion based DCO system traffic ($90\%$ of all traffic). The measured $-3.57\%$ CPM significant drop when compared to the CTR baseline may be explained by similar reasons and by the fact that we actually measure also a CTR drop. It is concluded that in the current A/B test setting where the main ranking models are dominated by the conversion based DCO system traffic, we can not measure actual revenue lifts. As a matter of fact, the conversion based DCO system is expected to increase revenue proportionally to the CVR lift.
%\oren{Explain this better..}
This is since the expected revenue (ad score) for oCPC ads, which consist of the lion share of the targeted DCO traffic, is $\mathrm{pCTR}\cdot \mathrm{pCONV} \cdot \mathrm{tCPA} \approx \mathrm{pCVR}\cdot \mathrm{tCPA} \approx \mathrm{CVR} \cdot \mathrm{tCPA}$ (see Section \ref{sec:serving}) \alex{where does the approx equality $\mathrm{pCTR}\cdot \mathrm{pCONV} \cdot \mathrm{tCPA} \approx \mathrm{pCVR}\cdot \mathrm{tCPA}$ come from?}\oren{assuming post-click conversions we have (clicks/impressions) x (conversions/clicks) = conversions/impressions = CVR} assuming our main models are accurate.

We measure a huge CPA drop of $-34.6\%$ and $-30.1\%$, which means that advertisers allegedly pay much less for conversions. However, for oCPC ads this is actually an \textit{artifact} stemming from the settings of our A/B testing where all buckets share the same main models used for ad-ranking. When measuring CPA, the baselines actually ``suffer'' from the situation where delivery is determined by the main ranking models, which reflects the surge in the DCO ads' CVRs, but combinations are actually drawn uniformly at random (at the Uniform bucket) or according to their CTR (in the CTR bucket). Hence, in the baselines buckets advertisers wrongfully spend more for less conversions. As a matter of fact, assuming our main models are accurate, no CPA drop is expected for oCPC DCO ads.

We end this section by noting that the evaluation was dominated by a few active DCO ads with relatively high CVRs. Milder CVR lifts are expected for an evaluation done with larger and more typical DCO demand.
%\vspace{-0.3cm}

\section{Concluding remarks}\label{sec:Concluding remarks}
In this work we presented a simple yet practical post-auction two-stage solution to the conversion optimization DCO problem. During the first stage of serving, a regular Gemini native auction is performed and if a DCO ad wins the auction, the Serving system uses the combination distributions generated by our combination CVR prediction based DCO solution, to draw the rendered combination. Since the solution is based on predictions provided by an auxiliary prediction model, and not on event counting used by the CTR based DCO system of \cite{koren2020dynamic}, it is capable of following trends and mitigate data sparsity issues. After showing good online performance, significantly increasing the DCO ads' CVRs, the conversion DCO product was deployed and has been serving all Gemini native traffic since. 
We note that different approaches to the conversion based DCO were considered as well. The simplest approach would have been to consider each combination as a new native ad and add it to the ad inventory. This however, could increase in theory the ad inventory and main click and conversion model sizes by a factor of the average number of combinations (e.g., $27$ assuming $3$ attributes with $3$ assets each). Another approach considered is treating the different combinations as ad features and incorporating those into \offset models. This approach would reduce the QPR of the Serving system since we need to calculate pCTRs and pCONVs for each combination per DCO ad before conducting the auction, thereby increasing the SLA by a factor of the average number of combinations. The proposed solution, requires minor modification of the Serving system and is not claimed to be optimal, does not increase \offset click and conversion model sizes and only slightly impacts the Serving system SLA. Moreover, it is demonstrated to use the additional degree-of-freedom provided by having multiple assets per attribute to increase CVR considerably over the baselines.

Future work may include improving the auxiliary prediction model accuracy, experimenting with additional user features and refining traffic segmentation. Another direction, is experimenting with a CTR prediction based DCO model. Once proven, we plan to combine both solutions and provide a unified CTR/CVR prediction based DCO solution to advertisers. 

\balance
\bibliographystyle{plain}
%\bibliography{references}

\begin{thebibliography}{10}

\bibitem{agarwal2009online}
D.~Agarwal, {B.C.}~Chen, P.~Elango, N.~Motgi, S.T.~Park, R.~Ramakrishnan, S.~Roy, and J.~Zachariah.
\newblock Online models for content optimization.
\newblock In {\em Proc. {NIPS'2009}}.

\bibitem{agrawal2012analysis}
S.~Agrawal and N.~Goyal.
\newblock Analysis of {Thompson} sampling for the multi-armed bandit problem.
\newblock In {\em Proc. Conference on Learning Theory}, 2012.

\bibitem{aharon2013off}
M.~Aharon, N.~Aizenberg, E.~Bortnikov, R.~Lempel, R.~Adadi,
  T.~ Benyamini, L.~Levin, R.~Roth, and O.~Serfaty.
\newblock \offset: one-pass factorization of feature sets for online recommendation in persistent cold start settings.
\newblock In {\em Proc. {RecSys'2013}}.

\bibitem{aharon2015serving}
M.~Aharon, A.~Kagian, Y.~Kaplan, R.~Nissim, and O.~Somekh.
\newblock Serving ads to" yahoo answers" occasional visitors.
\newblock In {\em Proc. {WWW'2015}}.

\bibitem{aharon2017adaptive}
M.~Aharon, A.~Kagian, and O.~Somekh.
\newblock Adaptive online hyper-parameters tuning for ad event-prediction models.
\newblock In {\em Proc. {WWW'2017}}.

\bibitem{aharon2019soft}
M.~Aharon, Y.~Kaplan, R.~Levy, O.~Somekh, A.~Blanc, N.~Eshel, A.~Shahar, A.~Singer, and A.~Zlotnik.
\newblock Soft frequency capping for improved ad click prediction in yahoo {Gemini} native.
\newblock In {\em Proc. {CIKM'2019}}.

\bibitem{aharon2019carousel}
M.~Aharon, O.~Somekh, A.~Shahar, A.~Singer, B.~Trayvas, H.~Vogel, and D.~Dobrev.
\newblock Carousel ads optimization in yahoo gemini native.
\newblock In {\em Proc. {KDD'2019}}.

\bibitem{arian2019feature}
M.~Arian, E.~Abutbul, M.~Aharon, Y.~Koren, O.~Somekh, and
  R.~Stram.
\newblock Feature enhancement via user similarities networks for improved click prediction in yahoo {Gemini} native.
\newblock In {\em Proc. {CIKM'2019}}.

\bibitem{audibert2010best}
J.Y.~Audibert and S.~Bubeck.
\newblock Best arm identification in multi-armed bandits.
\newblock In {\em Proc. {COLT'2010}}.

\bibitem{doi:10.1137/1.9781611974997}
A.~Beck.
\newblock {\em First-Order Methods in Optimization}.
\newblock Society for Industrial and Applied Mathematics, Philadelphia, PA, 2017.

\bibitem{besbes2014stochastic}
O.~Besbes, Y.~Gur, and A.~Zeevi.
\newblock Stochastic multi-armed-bandit problem with non-stationary rewards.
\newblock In {\em Proc. {NIPS'2014}}.

\bibitem{bubeck2012regret}
S.~Bubeck, N. Cesa-Bianchi, et~al.
\newblock Regret analysis of stochastic and nonstochastic multi-armed bandit problems.
\newblock {\em Foundations and Trends{\textregistered} in Machine Learning}, 5(1):1--122, 2012.

\bibitem{chakrabarti2009mortal}
D.~Chakrabarti, R.~Kumar, F.~Radlinski, and E.~Upfal.
\newblock Mortal multi-armed bandits.
\newblock In {\em Proc. {NIPS'2009}}.

\bibitem{chapelle2011empirical}
O.~Chapelle and L.~Li.
\newblock An empirical evaluation of thompson sampling.
\newblock In {\em Proc. {NIPS'2011}}.

\bibitem{duchi2011adaptive}
J.~Duchi, E.~Hazan, and Y.~Singer.
\newblock Adaptive subgradient methods for online learning and stochastic optimization.
\newblock {\em The Journal of Machine Learning Research}, pages 2121--2159, 2011.

\Comment{
\bibitem{edelman2007internet}
B.~Edelman, M.~Ostrovsky, and M.~Schwarz.
\newblock Internet advertising and the generalized second-price auction: Selling billions of dollars worth of keywords.
\newblock {\em The American economic review}, 97(1):242--259, 2007.}

\Comment{\bibitem{even2002pac}
E.~Even-Dar, S.~Mannor, and Y.~Mansour.
\newblock {PAC} bounds for multi-armed bandit and markov decision processes.
\newblock In {\em Proc. International Conference on Computational Learning Theory}, 2002.}

\bibitem{hillel2013distributed}
E.~Hillel, Z.~Karnin, T.~Koren, R.~Lempel, and O.~Somekh.
\newblock Distributed exploration in multi-armed bandits.
\newblock In {\em Proc. {NIPS'2013}}.

\bibitem{hofmann2013balancing}
K.~Hofmann, S.~Whiteson, and M.~de~Rijke.
\newblock Balancing exploration and exploitation in listwise and pairwise online learning to rank for information retrieval.
\newblock {\em Information Retrieval}, 16(1):63--90, 2013.

\bibitem{juan2017field}
Y.~Juan, D.~Lefortier, and O.~Chapelle.
\newblock Field-aware factorization machines in a real-world online advertising system.
\newblock In {\em Proc. {WWW'2017}}.

\bibitem{juan2016field}
Y.~Juan, Y.~Zhuang, W.S.~Chin, and C.J. Lin.
\newblock Field-aware factorization machines for ctr prediction.
\newblock In {\em Proc. {RecSys'2016}}.

\bibitem{kaplan2021dynamic}
Y.~Kaplan, Y.~Koren, R.~Leibovits, and O.~Somekh.
\newblock Dynamic Length Factorization Machines for CTR Prediction
\newblock In {\em Proc. {IeeeBigData'2021}}.

\bibitem{karnin2013almost}
Z.~Karnin, T.~Koren, and O.~Somekh.
\newblock Almost optimal exploration in multi-armed bandits.
\newblock In {\em Proc. {ICML'2013}}.

\bibitem{koren2020dynamic}
Y.~Koren, O.~Somekh, A.~Shahar, A.~Itzhaki, T.~Cohen, M.~Krasteva, and T.~Shadi.
\newblock Dynamic creative optimization in verizon media native advertising.
\newblock In {\em Proc. {IeeeBigData'2020}}.

\bibitem{lempel2012hierarchical}
R.~Lempel, R.~Barenboim, E.~Bortnikov, N.~Golbandi, A.~Kagian,
  L.~Katzir, H.~Makabee, S.~Roy, and O.~Somekh.
\newblock Hierarchical composable optimization of web pages.
\newblock In {\em Proc.{WWW'2012}}.

\bibitem{li2010contextual}
L.~Li, W.~Chu, J.~Langford, and R.E.~Schapire.
\newblock A contextual-bandit approach to personalized news article recommendation.
\newblock In {\em Proc. {WWW'2010}}.

\bibitem{rendle2010factorization}
S. Rendle.
\newblock Factorization machines.
\newblock In {\em Proc. IEEE International Conference on Data Mining}, 2010.

\bibitem{thompson2018now}
D.J.~Russo, B. Roy, A.~Kazerouni, I.~Osband, and Z.~Wen.
\newblock A tutorial on Thompson sampling.
\newblock {\em Foundations and Trends® in Machine Learning}, 11(1):1--96, 2018.

\bibitem{silberstein2020ad}
N.~Silberstein, O.~Somekh, Y.~Koren, M.~Aharon, D.~Porat, A.~Shahar, and T.~Wu.
\newblock Ad close mitigation for improved user experience in native advertisements.
\newblock In {\em Proc. {WSDM'2020}}.

\bibitem{slivkins2013ranked}
A.~Slivkins, F.~Radlinski, and S.~Gollapudi.
\newblock Ranked bandits in metric spaces: learning diverse rankings over large document collections.
\newblock {\em Journal of Machine Learning Research}, 14(Feb):399--436, 2013.

\bibitem{slivkins2008adapting}
A.~Slivkins and E.~Upfal.
\newblock Adapting to a changing environment: the brownian restless bandits.
\newblock In {\em Proc. {COLT'2008}}.

\bibitem{su2020attention}
Y.~Su, L.~Zhang, Q.~Dai, B.~Zhang, J.~Yan, D.~Wang, Y.~Bao,
  S.~Xu, Y.~He, and W.~Yan.
\newblock An attention-based model for conversion rate prediction with delayed feedback via post-click calibration.
\newblock In {\em Proc. {IJCAI'2020}}.

\bibitem{sutton2018reinforcement}
R.S.~Sutton and A.G.~Barto.
\newblock {\em Reinforcement learning: An introduction}.
\newblock MIT press, 2018.

\bibitem{wen2020entire}
H.~Wen, J.~Zhang, Y.~Wang, F.~Lv, W.~Bao, Q.~Lin, and K.~Yang.
\newblock Entire space multi-task modeling via post-click behavior decomposition for conversion rate prediction.
\newblock In {\em Proc. {SIGIR'2020}}.

\bibitem{wojdynski2016native}
B.W.~Wojdynski and G.J.~Golan.
\newblock Native advertising and the future of mass communication, 2016.

\end{thebibliography}

\end{document}